# The Case for RodentStore, an Adaptive, Declarative Storage System

Philippe Cudre-Mauroux, Eugene Wu, Samuel Madden
MIT CSAIL
{pcm,sirrice,madden}@csail.mit.edu

## ABSTRACT

Recent excitement in the database community surrounding new applications—analytic, scientific, graph, geospatial, etc.—has led to an explosion in research on database storage systems. New storage systems are vital to the database community, as they are at the heart of making database systems perform well in new application domains. Unfortunately, each such system also represents a substantial engineering effort including a great deal of duplication of mechanisms for features such as transactions and caching. In this paper, we make the case for RodentStore, an adaptive and declarative storage system providing a high-level interface for describing the physical representation of data. Specifically, RodentStore uses a declarative *storage algebra* whereby administrators (or database design tools) specify how a logical schema should be grouped into collections of rows, columns, and/or arrays, and the order in which those groups should be laid out on disk. We describe the key operators and types of our algebra, outline the general architecture of RodentStore, which interprets algebraic expressions to generate a physical representation of the data, and describe the interface between RodentStore and other parts of a database system, such as the query optimizer and executor. We provide a case study of the potential use of RodentStore in representing dense geospatial data collected from a mobile sensor network, showing the ease with which different storage layouts can be expressed using some of our algebraic constructs and the potential performance gains that a RodentStore-built storage system can offer.

## Categories and Subject Descriptors

H.2.4 [**Database Management**]: Systems—*relational databases*; H.2.8 [**Database Management**]: Database Applications—*Scientific databases*; H.2.8 [**Database Management**]: Database Applications—*spatial databases and GIS*

## 1. INTRODUCTION

> The rat ... is one of nature's most resourceful, adaptive creatures. ... The average rat can wriggle through a hole the size of a quarter, scale a brick wall, tread water for several days, gnaw through lead pipes and cinder blocks, survive a five-story fall, survive being flushed down a toilet, and even enter a building through the same route.
> -Brian Handwerk, National Geographic News, "Canada Province Rat-Free for 50 Years", March 31, 2003.

New database storage systems—based on something other than pure, row-oriented layouts—have been a popular topic in the database research literature. The recent resurgence in storage system design is motivated by growing commercial markets (such as OLAP) and new applications (e.g., scientific and geospatial settings). The Decomposition Storage Model (DSM) [13], for instance, considers a separate relation for each attribute and led to the development of highly-efficient column-store databases [20, 23], which can be orders of magnitude faster than traditional databases for OLAP workload queries. Refinements of the original decomposed-storage model also include the fractured mirrors approach [21], and PAX [6], which stores partitions of individual attributes in mini-pages. Beyond representations for traditional relational data, a growing number of storage systems [11, 14, 19, 24, 25] and languages [15, 18] focusing on arrays, geo-spatial data, "big science", and biology have been proposed.

Storage system research is vital to the database community, as it is at the heart of making database systems perform well in new application domains. Unfortunately, research systems are often far from usable by outside users, as building a robust storage engine requires a great deal of supporting code, including transaction, lock, and memory management facilities. Such features must be replicated in each stand-alone storage system. To address this challenge, we are building RodentStore, an open-source *adaptive storage system* that supports a variety of physical storage alternatives and can be used as the backing store for many different relational and array structured databases, rather than reimplementing a storage system for each new domain. By adaptive, we mean that RodentStore:

1. Can represent a variety of different physical designs, including, rows, columns, arrays, etc., and can store different tables using different physical representations. Furthermore, a single table can be stored using several different schemes (e.g., a mix of rows and columns.)

2. Provides a straightforward mechanism, the *storage algebra*, whereby a database administrator or a database design tool can specify the physical layout of data in the database, and can easily transform the physical representation of data. In this algebra, expressions represent transformations of the natural row-based database layout that would be derived from a logical database design. Operators specify how to reorder, group, transpose, merge, and compress rows and columns of the table. Operators are composed into nested expressions so that, for example, a table can be split into a collection of columns, each of which can be compressed or chunked in different ways.

This algebra-based approach makes it easy to explore different physical design alternatives, enables simple transformations between storage representations, and provides a layer upon which automated database design tools can be built. Furthermore, it makes it possible to explore combinations of storage schemas that haven't been considered before.

For example, given a database of sales records of the form:

```
N = (zipcode:z, year:y, month:m, day:d,
  customerid:c, productid:p ... )
```

The algebraic expression:

$$zorder(grid_{[y, z]}(N))$$

would repartition (or $grid$) the tuples into a matrix where years ($y$) are on the X axis and zipcodes ($z$) on the Y axis. Cells would be stored on disk using a space filling curve ($zorder$), so that nearby zipcodes or years are co-located.

Several detailed storage models were proposed in the mid-eighties for the study of database performance [8] or the unambiguous description of physical structures [7]. Contrary to those models, we focus our efforts on a higher-level and declarative description of the storage layout. We do not provide exhaustive descriptions of the storage structures, but rather focus on expressing the decomposition of logical tables into lists of relatively large chunks of data in order to model the recent storage placement schemes described above. In addition to the simple constructs supported by previous languages, however, we model important aspects of modern storage systems, such as compression schemes, multidimensional arrays, or nested structures.

Our approach is different from that taken in most current commercial systems, where a more conservative approach is called for. Major commercial database systems, for example, all offer tools to automatically select auxiliary data structures like indices and materialized views [4, 26] or partitioning techniques [5, 17, 21] in order to compensate for suboptimal physical data placements. These tools represent a significant first-step towards adaptive storage systems, but are limited in their scope, focusing typically on only one database system and one storage layout. Similarly, materialized views, though a valuable physical tuning tool, only allow database administrators to create tabular structures. Storing data as arrays, expressing unusual orderings (like z-order) or compression methods, and creating nested storage structures is not possible.

In the rest of this paper, we make the case for RodentStore, focusing on several key aspects of its design, including: the formulation of the storage algebra for describing equivalent physical representations of a given logical schema; a sketch of our proposed RodentStore storage engine implementation; a description of how the algebra can be used by a query optimizer to enumerate different storage policies; and a case study with geospatial data, showing that appropriate, high-level physical reorganization can offer orders-of-magnitude performance gains.

There are a few issues that we explicitly *don't* discuss in this paper, even though they will be included in our initial RodentStore prototype. The first is indexing—RodentStore will include both B+Trees as well as a variety of geo-spatial indices, but we don't anticipate innovating in this regard so we don't discuss it here. The second is distribution; we think it is essential that RodentStore supports multi-site distribution but focus our discussion here on the storage of tables on a single node.

## 2. ADAPTIVE STORAGE ARCHITECTURE

Figure 1 shows the basic architecture of RodentStore. Using a logical schema and workload as input, the database administrator or the optimizer specify the algebra that defines the physical arrangement of the storage system. The algebra interpreter compiles this algebra into a physical storage plan (or a plan that transforms the current representation into the new representation.) Commands to create and transform this representation are passed to the storage backend. The backend stores data according to this representation and implements methods for manipulating and retrieving data stored on disk.

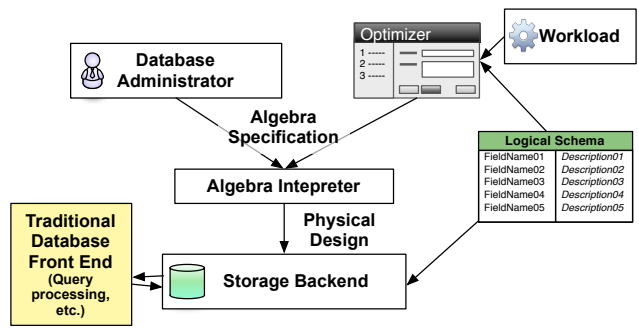

Figure 1: The RodentStore architecture.

We describe the storage algebra more detail in Section 3. We then discuss the storage backend, including its interface to the algebra interpreter and front end. Finally, we describe our initial ideas regarding the algebra optimizer in Section 5.

We do not discuss the front end in detail here; it may be a SQL database, an array oriented system, or any other interface (for example, object-relational mapping systems like Ruby-on-Rails and Django could use RodentStore as a backend.) We note that to get optimal performance out of a storage backend, the front end may need to be optimized; for example, in our work on column-stores, we showed that many optimizations inside the query processing are possible in addition to the advantages of column-oriented layout and compression [3].

## 3. STORAGE ALGEBRA

In the following, we introduce the basic principles and operators of our storage algebra, which is used to formulate storage variants for a given logical schema. The basic idea of our algebra is to specify physical layout transformations from the canonical representation of row-oriented tables. The storage algebra supports a set of declarative constructs to describe, combine, or modify arbitrary layouts defined as nested structures. These declarations can then be compiled and efficiently executed using various optimization strategies. We give a short overview of the main features of our language below.

### 3.1 Physical Layout Dimensions

Laying out physical objects on disk involves optimizations in a number of dimensions, which we briefly review here.

**Data Co-location and Isolation** regroups co-accessed data in the same vicinity on the storage medium. Co-location of data helps to minimize random accesses and to avoid cache misses when fetching data to answer a query. Isolation considers dedicated storage space to pull apart subsets of data, for example for isolating elements that are frequently updated.

**Data Reduction** aims at maximizing the effective data throughput by minimizing the space needed to store data. It encompasses both dense-packing, which strives to minimize dead space and superfluous metadata like tuple headers, and data compression schemes.

**Data Reordering** can be used to enforce specific arrangements of successive data values on disk. It can be used to channel data according to a given access pattern or sorting criterion to avoid data reshuffling and repeated data scans. Data ordering can also be exploited to efficiently navigate through regular data structures like ordered lists of fixed-size elements or arrays, by direct-offsetting to particular values.

We divide our discussion of operations in our algebra according to these dimensions.

## 3.2 Data Model

We consider logical representations of databases comprising collections of tables. Each table consists of a set of records following the same schema. More formally, we say that a database $D$ is a set of tables $T$ such that each table $t \in T$ contains a set of $m_t$ records $r_1, \ldots, r_{m_t}$, where each record contains $n$ elements (e.g., fields) $e_1, \ldots, e_n$.

The storage algebra manipulates lists of elements through successive nestings and transformations. Each element is associated with one of the data types supported by the algebra:

$$\tau := int \mid float \mid string \mid \ldots \mid l : \tau \mid [\tau_1, \ldots, \tau_n]$$

where the first few entries $int \mid float \mid string \mid \ldots$ represent a collection of data types, either of fixed or variable size, commonly supported by database systems. $l : \tau$ names an element by associating a literal to a given type, while the nesting clause $[\tau_1, \ldots, \tau_n]$ allows the creation of arbitrary *nestings* of types.

## 3.3 Nestings

Nesting clauses $[\cdot]$ are the primary focus of the storage algebra; they regroup lists of objects $[\tau_1, \ldots, \tau_n]$ that are to be stored together. Nestings can be defined in a straightforward manner by enumerating lists of elements, e.g.,

$$N_0 = [[1, 2, 3], [12, 13, 14]]$$

defines a list of two elements, each containing a list of three integers.

In addition, the storage algebra supports a more powerful way of defining nestings through list comprehensions [10, 27]. List comprehensions are akin to well-known set definitions and provide a compact and expressive way of defining lists. They are used by the storage algebra to declare new nestings from existing nestings. In the context of the storage algebra, simple list comprehensions take the following generic form:

$$e(v) \mid \backslash v \leftarrow N, C$$

where $\backslash v \leftarrow N$ is a *generator* binding a variable $\backslash v$ to the successive elements of an existing nesting $N$, $C$ is a condition, and $e$ represents the elements of the resulting nesting written in terms of the variable $v$. We allow several generators and conditions per comprehension.

As an example, consider a simple table $T = [[Zip : int, Area : int, Addr : string]]$ storing zip and area codes associated with addresses. By default, suppose that $T$ stores its records in an arbitrary order using a row-major representation. We can express this representation explicitly using a list comprehension as follows:

$$\mathbf{N_r} = [[r.Zip, r.Area, r.Addr] \mid \backslash r \leftarrow T]$$

$$\equiv \begin{bmatrix} [zip_1 & Area_1 & Addr_1] \\ \vdots & \vdots & \vdots \\ [zip_m & Area_m & Addr_m] \end{bmatrix}$$

We can express a different, column-major storage layout for the same logical schema as follows:

$$\mathbf{N_c} = [[r.Zip \mid \backslash r \leftarrow T], [r.Area \mid \backslash r \leftarrow T], [r.Addr \mid \backslash r \leftarrow T]]$$

$$\equiv \begin{bmatrix} \begin{bmatrix} zip_1 \\ \vdots \\ zip_m \end{bmatrix} \begin{bmatrix} Area_1 \\ \vdots \\ Area_m \end{bmatrix} \begin{bmatrix} Addr_1 \\ \vdots \\ Addr_m \end{bmatrix} \end{bmatrix}$$

In the storage algebra, conditions $C$ represent either boolean valued expressions or clauses [16] that limit the number of elements returned (*limit*), order elements (*orderby*), regroup (*groupby*) or partition (*partitionby*) elements into sub-nestings. The algebra also includes helper functions, e.g., to return the position of an element (*pos()*) or the number of elements contained in a nesting (*count()*). The following comprehension returns, for example, a sorted list of zip codes associated with a given area code:

$$N_z = [r.Zip \mid \backslash r \leftarrow T, \ r.Area = 617, \ orderby \ r.Zip \ ASC]$$

List comprehensions can be physically executed using a number of implementations, including recursive functions and nested iterators (see Section 4).

## 3.4 Physical Representation

All the storage schemes discussed in the introduction (e.g., column-stores, DSM, PAX, etc.) can be seen as laying out data in hierarchically organized chunks on disk. RodentStore takes advantage of nesting clauses to describe such storage schemes in a generic manner. Nesting clauses represent a natural way to describe stored data, as they consider ordered lists of elements that can be nested arbitrarily. In addition to relations, nestings can also naturally support time-series values and multidimensional objects, as we describe in Section 3.6 below. RodentStore lays out data on disk by flattening the nesting clauses given as input. The *physical representation* of a nesting gives the order with which the data is to be written:

**Physical Representation** The physical representation $\phi(N)$ of a nesting $N$ is obtained by recursively enumerating all its entries starting from the leftmost entry. Hence, the physical representation $\phi(N)$ of a nesting $N$ can be represented as a list of entries: $\phi(N) = [e_1, \ldots, e_N]$

In Section 4, we discuss algorithms for generating such representations.

## 3.5 Transforms

The storage algebra includes a series of functions (a.k.a. *transforms*) that capture various dimensions of physical layout optimization. We briefly describe some of those transforms below.

### 3.5.1 Data Co-location & Isolation

The *project* transform is an important construct that isolates an element or a series of elements from a nesting:

$$project_{[A_i, \ldots, A_j]}(N) \equiv [[r.A_i, \ldots, r.A_j] \mid \backslash r \leftarrow N]$$

The reciprocal transform, $append([e_1, \ldots, e_m], N)$, attaches elements to the tuples in $N$. $select_{C(A)}(N)$ returns all elements $A$ in $N$ satisfying the condition $C$. Horizontal partitioning, finally, is supported through $partition_C(N)$, which subdivides all first-level nestings of $N$ using a condition $C$.

### 3.5.2 Data Reduction

For each value $a$ of some (set of) attribute(s) $A$, the *fold* operation nests values $b$ of attribute(s) $B$ that co-occur with $a$:

$$fold_{B,A}(N) \equiv [r.A, [r'.B \mid \backslash r' \leftarrow N, r.A = r'.A] \mid \backslash r \leftarrow N]$$

This transform can be reversed by *unfold*. *fold*, can, for example, be used to group of a list of zip codes and addresses with each area code in our previous example:

$$\begin{bmatrix} Area_1, [[Zip_{11}, Addr_{11}], \ldots [Zip_{1n}, Addr_{1n}]] \\ Area_2, [[Zip_{21}, Addr_{21}], \ldots [Zip_{2n}, Addr_{2n}]] \\ \ldots \end{bmatrix}$$

*fold* is most useful when run over denormalized data generated by the $prejoin$ transform, since prejoined data is likely to include many repeated values:

$$prejoin_{joinatt}(N_1, N_2) \equiv$$
$$[[r_1, r_2] \mid \backslash r_1 \leftarrow N_1, \backslash r_2 \leftarrow N_2, r_1.joinatt = r_2.joinatt]$$

In addition, the storage algebra supports a wide range of compression schemes by producing nestings through user-defined functions. As an example, here is a delta compression scheme, which can be used to compress time series or ordered values by considering the differences between subsequent elements:

$$\Delta(N) \equiv [a - b$$
$$\mid [a, b] \leftarrow [N, [0, \ n \mid \backslash n \leftarrow N, limit \ count(N) - 1]]].$$

### 3.5.3 Data Reordering

Reordering of data is supported by the *orderby* clause. *orderby* can reorder elements according to several attributes (as in SQL), and can also be used to reorder complex sub-nestings. $zorder(N)$, for example, takes into account the position of both first-order and second-order nested elements in $N$ to rearrange the elements according to a z-order traversal of the structure:

$$zorder(N) \equiv [r' \mid \backslash r \leftarrow N, \backslash r' \leftarrow r, \ r' \ orderby$$
$$interleave(bin(pos(r)), bin(pos(r'))) \ ASC]$$

where

$$interleave(A, B) \equiv [a, b \mid [\backslash a, \backslash b] \leftarrow [A, B]]$$

is used to interleave the bits of the binary representation (*bin*) of the position of the elements to produce the proper z-order.

## 3.6 Arrays

As can be seen by the previous transform, multidimensional data such as arrays can be naturally supported by our algebra by specifying several levels of nestings. For instance,

$$N_m = [[1, 2, 3], [4, 5, 6]]$$

represents a 3 x 2 matrix stored in a row-major fashion. Many common multidimensional operations, such as matrix transpositions, are easy to express using our formalism:

$$transpose(N) \equiv$$
$$[[a_i, \ldots, n_i] \mid [\backslash a, \ldots, \backslash n] \leftarrow N, \backslash a_i \leftarrow a, \ldots, \backslash n_i \leftarrow n].$$

For example, $transpose(N_m)$ run on the $N_m$ matrix would produce $[[1, 4], [2, 5], [3, 6]]$. The storage algebra defines several transforms to chunk arrays for storage purposes [22] and to shift from unidimensional to multidimensional representations. The $grid$ transform, for instance, creates a $n$-dimensional array from a list of tuples by repartitioning the tuples along $n$ discretized dimensions:

$$grid_{[A_1,\ldots,A_n],[stride_1,\ldots,stride_n]}(N) \equiv$$
$$[r \mid \backslash r \leftarrow N, partition by \ r.A_1 \ stride_1, \ldots, r.A_n \ stride_n]$$

Section 6 shows how we can use this transform to rearrange lists of tuples on a 2D plane.

This completes the summary of the storage algebra. We showed how its expressions can capture a range of storage options, including storing arrays, converting between columns and rows, and decomposing and grouping relations in a variety of ways.

## 4. RODENTSTORE IMPLEMENTATION

In this section, we focus on the API for data access and cost estimation (used, for example, by the query processor front-end) and the translation of storage algebra expressions into physical, on-disk structures.

## 4.1 API

As a storage system, RodentStore exposes a simple API that allows higher layers to iterate through tuples of a table and to estimate the cost of various access paths into the storage system. In that sense, it is similar to the access methods exposed by purely relational storage systems.

RodentStore provides the following methods:
1. `scan(table, [fieldlist, predicate, order])`: Scans an entire relation, with optional projection, range predicate, and sort order.
2. `getElement(table, [fieldlist,] index)`: Returns the element at position `index` of the relation; if the table is stored as an array, `index` may be multidimensional.
3. `next(table, [order])`: called after `getElement`; returns the next element in the (optional) order `order`, or in whatever the default order of the table is.
4. `scan_cost(table, [fieldlist, predicate, order])`: Returns the estimated cost, in milliseconds, of the specified scan operation.
5. `getElement_cost(table, [fieldlist,] index)`: Returns the estimated cost, in milliseconds, of the specified `getElement` operation.
6. `order_list(table)`: Retrieves a list of sort orders for which the current storage organization is "efficient".

To implement the scan methods, RodentStore maintains cursors on each constituent object of a relation, and attempts to store and walk each object in the same order whenever possible. For example, if a table is stored as a collection of vertical partitions and no ordering clause is specified on those vertical partition, then these partitions will be stored and traversed in the same order. (If the storage algebra specifies the vertical partitions should be stored in different orders, then RodentStore may have to re-sort the data before returning it.) As another example, if a particular attribute is nested inside of a parent tuple, each inner value is "unnested" by merging with the parent and outputting the entire tuple. Exploring variants of the API that, for example, emit blocks of tuples that are nested or run-length compressed (as in Abadi *et al.* [1]) is an interesting research direction.

The `scan_cost`, `getElement_cost`, and `order_list` methods are designed to be used by the query optimizer of a database system sitting on top of RodentStore; additional cost methods may be necessary as we further develop RodentStore. We also anticipate adding methods to extend RodentStore with new storage transforms; these will require additional methods for costing and possibly for iterating through the data.

## 4.2 Physical Layout

The algebra interpreter's job is to translate storage algebra expressions into on-disk structures. This problem is interesting for several reasons. First, translating a nested algebra expression with several array completions in it is analogous in many ways to database join evaluation. As with joins, the simplest way to evaluate an expression is through nested *for* loops. For example, consider the *fold* expression:

$$fold_{B,A}(N) \equiv [r.A, [r'.B \mid \backslash r' \leftarrow N, r.A = r'.A] \mid \backslash r \leftarrow N]$$

This can be "rendered" on disk using two *for* loops, as shown in Algorithm 1.

```
/* outerList records values from the outer
   we've already seen                         */
outerList = []
foreach r ∈ N do
    if r.A ∈ outerList then
        continue
    end
    /* innerList records values nested in
       outer tuple                            */
    innerList = []
    foreach r' ∈ N do
        if r'.A == r.A then
            innerList.append(r'.B)
        end
    end
    outerList.append(r.A)
    writeTuple([r.A, innerList])
end
```
**Algorithm 1**: Nested *for* loops implementing the *fold* transform.

Note, however, that rather than using nested *for* loops, a hash-join like algorithm could be used, where a first pass through $N$ builds a hash table on $r.B$, and then a second pass writes each $r.A$ along with the $r.B$ values that match it (which can be found via a single hash-table lookup.)

Similar rendering algorithms are needed for each of the transforms described above. For example, the $grid$ operator will require applying a group-by like operation to assemble the values in each cell of the grid, following by writing out the cells in the appropriate order.

The second interesting issue with storage rendering is that the storage algebra is "declarative", in the sense that it leaves many things unspecified. This means that there are many layout alternatives that RodentStore may consider. For example, for a given object (e.g., a column), in what order should the records in that object be written (assuming no ordering is explicitly specified)? Absent any kind of information about a "good" order (e.g., from a supplied workload), then at a minimum the storage layout algorithm should try to store different objects from the same table in the same order. This is important so that when the query processor iterates through tuples of a relation, the storage manager doesn't have to reorder those objects or maintain additional information about which objects are in a particular tuple.

There are a number of other questions along these lines that a layout engine can consider; for example: should objects be dense-packed or should additional free space be left for inserts? For arrays, should variable length fields be stored "out of band", so that direct offsetting can be used to lookup specific elements in the array? When multiple disks or multiple nodes in a cluster are available, how should objects be partitioned across those physical storage entities? What is the appropriate disk page size to use? What should the system do to adapt to storage on Flash or in main-memory (RAM-based) databases?

As the above list of question shows, there are a variety of interesting research issues related to storage layout that we plan to tackle as we build the RodentStore system.

## 5. STORAGE DESIGN OPTIMIZER

In many cases, we anticipate that the database administrator will want to manually specify the storage algebra for his or her database. However, we also plan to build a storage design optimizer in RodentStore, which takes as input a relational schema and a workload of SQL queries and outputs a recommended storage representation. Note that this is different from the layout optimizations presented in the previous section—our goal here is to choose the best algebraic expression, whereas the goal in the previous section was to choose the best physical layout for a given algebraic expression.

We anticipate this optimizer working similarly to modern physical design recommendation tools [9, 12]. Specifically, we plan to use a cost-based optimization method, which uses a cost model to estimate the cost of running the supplied workload against a series of candidate physical designs. The optimizer then searches through the space of possible designs and returns the one that minimizes the sum of costs of queries in the workload.

We do not anticipate innovating on the cost model; our initial plans are for it to count bytes of I/O as well as disk seeks, using the cost functions exposed by the RodentStore storage layer. We will ignore CPU costs unless we find that operations like decompression prove to contribute significantly to our overall runtime. Previous work [1] suggests that even heavyweight schemes like Lempel-Ziv offer greater time savings as a result of reduced I/O than they cost in terms of increased decompression time.

Plan enumeration is considerably trickier. Most of the above transformations lead to an exponential number of physical designs. For example, if there are $n$ columns in a table, there are $2^n$ ways to co-locate that table's columns. Similarly, a table with $n$ columns can be gridded into two dimensions in $O(n^2)$ ways, and in multiple dimensions in $O(2^n)$ ways. For this reason, we anticipate heavy reliance on heuristic search algorithms. For example, to find the best gridding, we could use gradient descent or simulated annealing to add dimensions until a low cost dimensionalization is achieved. Similarly, for partitioning, we anticipate taking advantage of previous works (e.g., [5]) to rapidly identify promising groups of columns that could get co-located. Exploring the ordering in which different transforms should be applied as well as the effectiveness of these various algorithms will form a significant part of the research in the RodentStore project.

Deciding what to do when a new physical design is created is a challenge; we plan to investigate a range of options. One choice is to *eagerly reorganize*, where every object with a new design is rewritten immediately. Another option is to *reorganize only new data*, leaving old data as it was. This is obviously much less expensive than eager reorganization, but may result in poor performance if the workload changes dramatically. It also complicates access method implementations, as new and old data must be merged; this is especially true if the query processor expects to receive results in a particular order, as RodentStore must then buffer and reorder data on the fly. Thus, we anticipate that a third, *lazy reorganization* approach, where objects are rewritten in the background or when they are accessed, may be superior, as it avoids the up-front delay of eager reorganization as well as the access method complexities of reorganizing new data only.

## 6. CASE STUDY

To illustrate the potential of RodentStore, we manually implemented a few of the transforms over a collection of geospatial data. Our implementation does not support arbitrary storage algebra constructs and transformations, providing only the constructs needed for the case study. The current prototype emerged from the need to efficiently query the very large data sets produced by CarTel (http://cartel.csail.mit.edu/)—a car telematics infrastructure that has been used to collect hundred of thousands of motion traces from a fleet of cars in Boston. CarTel data is of particular interest in the context of RodentStore as it can be viewed as relations, time-series values or multidimensional arrays. The data is currently stored in a PostgreSQL database, whose performance is not satisfactory for several interactive applications we are currently developing—in particular, visualization applications that need to browse large numbers of such traces are very slow. We describe below some of the physical representations supported by our prototype and discuss their respective effectiveness at answering queries over this data.

We focus on a relation storing raw GPS traces and on queries retrieving trajectory data given a geographic region expressed as a spatial rectangle. The logical schema of this data is:

$$Traces(int\ t, float\ lat, float\ lon, double\ ID, \ldots)$$

where $t$ is a timestamp attached to the observations, $lat$ and $lon$ are the latitude and longitude of the taxi, and $ID$ is a string uniquely identifying a taxi. There are a number of additional attributes for each reading that we omit. We first consider a classical, row-oriented physical layout:

$$N_1 = [[r.t, r.lat, r.lon, r.ID] \mid \backslash r \leftarrow Traces].$$

$N_1$ retrieves all tuples from $Traces$ and stores them contiguously on disk. Without any ordering or indices, spatial queries retrieving $(lat, lon)$ points within certain bounds are answered by performing a full table scan inspecting all tuples. A more efficient layout could order observations by time and regroup tuples by trajectory in order to drop all unnecessary attributes for this particular class of queries:

$$N_2 = [[r.lat, r.lon] \mid \backslash r \leftarrow N_1, orderby\ r.t, groupby\ r.ID]]$$

Since we in fact are processing two-dimensional queries and values, we repartition the data on disk taking advantage of a two-dimensional lattice:

$$N_3 = [grid_{cellHeight, cellWidth}(N_2)].$$

This nesting arranges the $lat$ and $lon$ values on a two-dimensional grid, groups them into cells of size $cellWidth \times cellHeight$, co-locates all values belonging to the same cell on disk, and creates a hash table that tracks the spatial boundaries of each cell. This representation is particularly efficient in our context as it allows the system to skip over all cells whose spatial boundaries do not overlap with the query. In addition, we reorder the cells on disk using a space-filling curve in order to minimize the disk seek times when retrieving spatially contiguous objects:

$$N_3' = [zorder(N_3)]$$

Finally, to compress the size of the scanned data even further, we apply delta compression on the latitude and longitude values:

$$N_4 = [\Delta[r.lat] \mid \backslash r.lat \leftarrow N_3',\ \Delta[r.lon] \mid \backslash r.lon \leftarrow N_3']$$

The rationale behind this nesting lies in the fact that cars move continuously by small increments on the spatial plane, and that it is more efficient to store these small increments rather than the absolute latitude and longitude values.

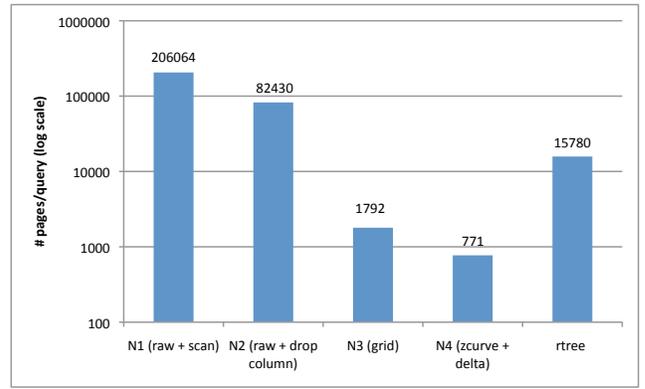

**Figure 2: Performance in terms of number of pages read by RodentStore prototype on geospatial trajectory data from the CarTel project.**

Figure 2 gives the average number of disk pages read for the various physical layouts described above. The results are averaged over 200 random geographical queries retrieving square regions covering 1% of the total area considered (the entire region covers the greater Boston area; as such, each grid cell is about $400\ m^2$.) We focused on a subset of the data centered around MIT and containing ten million observations (more than 200 MB of data, representing a few thousand trajectories), and a page size set to 1000 KB. Results using the original raw files ("raw+scan") are given for comparison purposes. The results show that data isolation ("drop column") and gridding ("grid") reduce the total number of pages that must be read by about two orders of magnitude versus a raw scan; the total query time (not shown) is also about one hundred times faster (a few 10s of milliseconds vs five seconds.) Using z-ordering reduces the number of disk seeks needed to fetch data in a given spatial region. Finally, delta compression ("zcurve + delta") reduces the database size, thus the I/O cost, even further. We also experimented with a relatively common approach to index spatial objects using a secondary R-Tree over the trajectories ("rtree"). Using an R-Tree is in our case suboptimal, given that our dense data set generated a high number of overlapping bounding boxes, each requiring a random I/O and containing a large number of observations.

## 7. CONCLUSIONS

In this paper, we made the case for RodentStore, an adaptive storage system that provides a high-level, declarative interface for describing and dynamically modifying physical data layouts. In light of recent excitement surrounding database storage schemes, a system like RodentStore is important because it substantially reduces the engineering efforts involved in building a storage manager for new types of data and in evaluating new layout mechanisms. RodentStore supports a wide range of physical structures, encompassing nested lists of tuples, time-series values, and multidimensional arrays. Our system can handle unusual storage schemes—such as attribute-dependent layouts for RDF data [2]—while still exposing logical tables or array schemas at the application layer. The general outline of the system given above opens the door to many research challenges. Our immediate efforts will focus on generic mechanisms for storing arbitrary nestings in a compact way, and on specific techniques for optimizing the transforms supported by the system.

## 8. ACKNOWLEDGMENTS

This research was supported by the NSF under grant number IIS-0704424.